# The First Light Curve Solutions of GW Leo and Period Study of Two Contact Binary Systems

**Atila Poro[1], Afshin Halavati[1,2], Elahe Lashgari[1], Ali Gardi[3], Keyvan GholizadehSoghar[3], Yasaman Dashti[1], Fatemeh Mohammadizadeh[1], Mahya Hedayatjoo[1]**

[1]The International Occultation Timing Association Middle East section, Iran, info@iota-me.com
[2]BiKaran Observatory, Kerman, Iran
[3]SabalanSky Astronomy Association, Ardebil, Iran

**Abstract**
We performed the first light curve analysis of GW Leo and a new ephemeris is obtained for QT Boo. In the present photometric study of two contact binary systems, we found that the period of these binary systems is decreasing at a rate of $\frac{dp}{dt} = -6.21 \times 10^{-3}\ days\ yr^{-1}$ for GW Leo, and $d\frac{dp}{dt} = -4.72 \times 10^{-3}\ days\ yr^{-1}$ for QT Boo, respectively. The light curve investigation also yields that the system GW Leo is a contact W UMa eclipsing binary with a photometric mass ratio of $q = 0.881 \pm 0.030$, a fillout factor of $f = 3\%$, and an inclination of $54.060 \pm 0.066\ deg$. Due to the O'Connell effect which is known as asymmetries in the light curves' maxima, a cold spot is employed along with the solution. We also calculate the distance of GW Leo from the distance modulus formula as $465.58 \pm 23$ pc, which is relatively close to the quantity measured by the Gaia DR2 using the binary systems' parallax. Moreover, the positions of their components on the H-R diagram are represented.

## 1. Introduction

EW-type eclipsing variables are considered as a subclass of W Ursae Majoris-type eclipsing binaries and have periods less than a day. EW eclipsing binaries may form through magnetic braking in which detached binary stars with short period lose angular momentum (Guinan & Bradstreet 1988). It is also discussed that the components in the W Uma type binary systems have nearly close temperatures because of their proximity (Paczyński et al. 2006).

We present the photometric observations of GW Leo and QT Boo as W UMa binary systems. The study includes the first light curve analysis of GW Leo and the determination of a new ephemeris for both systems. We find that systems selected in terms of magnitude and period are very suitable for investigation. GW Leo (GSC 00843-00262) which is located in the Leo constellation has a magnitude range from 13 and 13.134, and a period of 0.336157 days (Rinner 2003); and QT Boo (GSC 03486-01026) which is located in the Booties constellation has a magnitude range from 11.65 and 11.80, and a period of 0.319065 days (Khruslov 2008), respectively.

## 2. Observation and data reduction

GW Leo was observed with a 10-inch Schmidt Cassegrain telescope at the Bkaraan Observatory of Kerman, Iran (Long. 57 01 13 E, Lat. 30 16 55 N, Alt. 1764 m) in May 2020. During two nights, a total of 404 photometric data were obtained in the $V$ filter with an exposure time of 30 seconds for every single image. We used a Nikon D5300 DSLR Camera attached to the telescope.

QT Boo was another candidate in this study and it was observed in September 2020 at the Payame Noor University Observatory of Ardabil, Iran (Long. 48 17 42 E, Lat. 38 12 55 N, Alt. 1380 m). During two nights, a total of 159 images were taken in the standard filter of V. An 11-inch Schmidt Cassegrain telescope along with a Canon 600D DSLR Camera was employed to collect data. The exposure time for two times of minima was 30 seconds.

To perform differential photometry, we selected one reference star and one comparison star for each of the binary systems. The magnitude of the Variable, comparison, and reference stars, as well as their Right ascension and Declination, are available in Table 1.

**Table 1. Information derived from SIMBAD[1] and O-C Gateway[2] related to the variable, comparison, and reference stars.**

| Type | Star | Magnitude ($V$) | RA. (2000) | Dec. (2000) |
|---|---|---|---|---|
| Variable | GW Leo | 13.00-13.13 | 10 18 53.48 | +13 41 08.67 |
| Comparison | GSC 843-339 | 13.50 | 10 19 00.73 | +13 42 28.01 |
| Reference | GSC 843-419 | 13.00 | 10 18 34.04 | +13 44 47.83 |
| Variable | QT Boo | 11.65-11.80 | 15 35 10.98 | +49 47 44.05 |
| Comparison | GSC 3486-1435 | 13.40 | 15 35 31.87 | +49 42 41.14 |
| Reference | GSC 3486-1075 | 13.60 | 15 35 31.52 | +49 52 21.10 |

Using the Digital Single Lens Reflex camera (DSLR) has been currently a common method of photometry in the small observatories. The skill of the observer and the stage of data reduction are two main factors that determine the quality of data in this method (Davoudi et al. 2020).

Bias, dark, and flat-field frames were used to do data subtraction and correction. We employed AstroImageJ (AIJ) software to calibrate and perform differential aperture photometry (Collins et al. 2017). The minimum times were found through fitting the models to the light curves and existing minima based on Gaussian and Cauchy distributions. We then employed the Monte Carlo Markov Chain (MCMC) method to measure the amount of uncertainty related to each value (Poro et al. 2020). We also benefited from Python along with its PyMC3 package to execute the lines of code (Salvatier et al. 2016). Figure 1 shows the observed and synthetic light curves in the $V$ filter for GW Leo, and Figure 2 is the observed light curve of QT Boo.

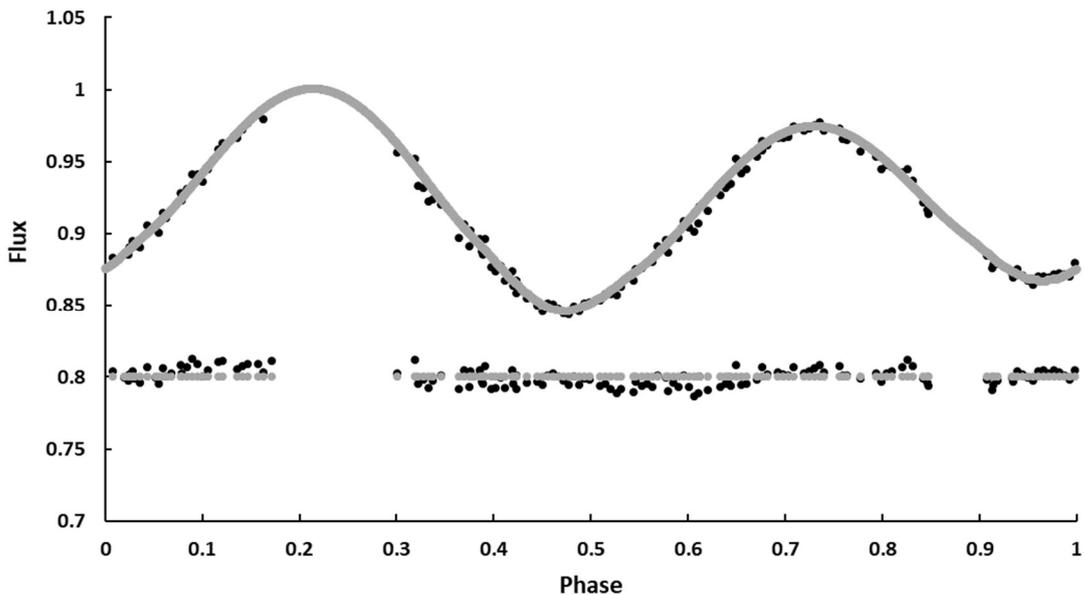

**Figure 1. The observed light curve of GW Leo and synthetic light curve in the $V$ filter and residuals are plotted.**

---

[1]http://simbad.u-strasbg.fr/simbad/
[2]http://var2.astro.cz/ocgate/

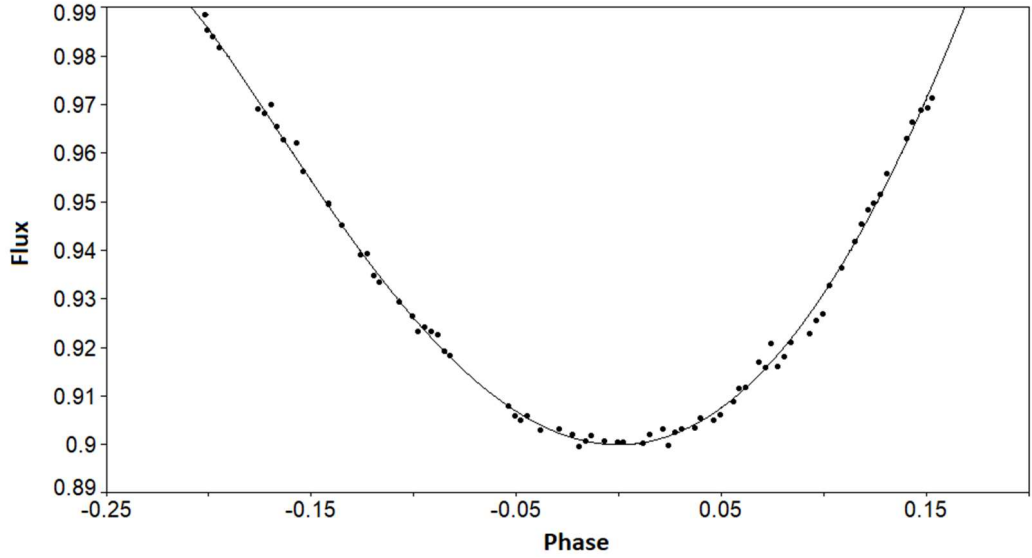

**Figure 2.** The observed light curve of QT Boo in the $V$ filter.

## 3. New ephemeris and period variations

Based on our observation, we determined two minimum times for each binary system and also collected 16 and 11 more minima for GW Leo and QT Boo respectively from the literature. They are all listed in the first column of Tables 2 and 3 in the Barycentric Julian Date in the Barycentric Dynamical Time (BJD$_{TDB}$) format, and their uncertainties appear in column 2. Epochs of these minima times and O-C values are in column 3 and 4 respectively; and the references of mid-eclipse times are shown in the last column.

Table 2. Available times of minima for GW Leo.

| BJD$_{TDB}$ | Error | Epoch | O-C | References |
|---|---|---|---|---|
| 2452721.5287 |  | 0 | 0 | Rinner 2003 |
| 2455642.4396 | 0.0007 | 8689 | 0.0427 | Hoňková 2013 |
| 2455642.4407 | 0.0006 | 8689 | 0.0438 | Hoňková 2013 |
| 2455992.3524 | 0.0009 | 9730 | 0.0161 | Hoňková 2013 |
| 2455992.3526 | 0.0005 | 9730 | 0.0163 | Hoňková 2013 |
| 2456011.3361 | 0.0007 | 9786.5 | 0.0069 | Hoňková 2013 |
| 2456011.3368 | 0.0004 | 9786.5 | 0.0076 | Hoňková 2013 |
| 2456330.4993 | 0.0007 | 10736 | -0.0110 | Hoňková 2013 |
| 2456330.5056 | 0.0007 | 10736 | -0.0047 | Hoňková 2013 |
| 2456330.5098 | 0.0011 | 10736 | -0.0005 | Hoňková 2013 |
| 2456670.6811 | 0.0004 | 11748 | -0.0201 | Hoňková 2015 |
| 2456670.6821 | 0.0010 | 11748 | -0.0191 | Hoňková 2015 |
| 2456745.4366 | 0.0008 | 11970.5 | -0.0595 | Hoňková 2015 |
| 2456745.4393 | 0.0005 | 11970.5 | -0.0568 | Hoňková 2015 |
| 2457798.5414 |  | 15103 | 0.0335 | BRNO 41[3] |
| 2457798.5425 |  | 15103 | 0.0346 | BRNO 41 |
| 2458987.2919 | 0.0011 | 18639.5 | -0.0352 | This study |
| 2458987.4599 | 0.0018 | 18640 | -0.0353 | This study |

---

[3]http://var2.astro.cz/brno/

Table 3. Available times of minima for QT Boo.

| $BJD_{TDB}$ | Error | Epoch | O-C | References |
|---|---|---|---|---|
| 2451402.5378 | | 0 | 0 | Khruslov 2008 |
| 2456368.4538 | 0.0008 | 15564 | -0.0117 | Hoňková 2013 |
| 2456745.4298 | 0.0052 | 16745.5 | -0.0110 | Hübscher 2015 |
| 2456745.5878 | 0.0027 | 16746 | -0.0125 | Hübscher 2015 |
| 2457027.6388 | | 17630 | -0.0150 | Jurysek 2017 |
| 2457100.3898 | 0.0022 | 17858 | -0.0108 | Hübscher 2016 |
| 2457100.5368 | 0.0042 | 17858.5 | -0.0233 | Hübscher 2016 |
| 2457124.7958 | | 17934.5 | -0.0132 | Nelson 2016 |
| 2457493.7888 | | 19091 | -0.0189 | Nelson 2017 |
| 2457807.9068 | 0.0030 | 20075.5 | -0.0204 | Nelson 2018 |
| 2457843.4818 | | 20187 | -0.0212 | Pagel 2018 |
| 2459104.4239 | 0.0040 | 24139 | -0.0239 | This study |
| 2459105.3807 | 0.0050 | 24142 | -0.0243 | This study |

The epochs and the O-C values were calculated according to the reference ephemeris for both binary systems (Table 4). We wrote a Python code based on the emcee package to fit all collected mid-eclipse times with a line as shown in Figures 3 and 4. We used the MCMC to determine a new linear ephemeris for the primary minimum, and therefore we have obtained the posterior probabilities for these parameters (Figure 5). Consequently, a new linear ephemeris related to the primary minimum for each system was determined and shown in Table 5. $E$ represents the number of cycles after the reference epoch, and the errors corresponding to the new mid-eclipse time, and the new period appeared in the brackets.

Table 4. The reference ephemeris of GW Leo and QT Boo.

| Binary system | Reference ephemeris | Reference |
|---|---|---|
| GW Leo | Min I. = $BJD_{TDB}$ 2452721.5287 + 0.336157 × $E$ | Rinner 2003 |
| QT Boo | Min I. = $BJD_{TDB}$ 2451402.5378 + 0.319065 × $E$ | Khruslov 2008 |

Table 5. The New ephemeris of GW Leo and QT Boo.

| Binary system | New ephemeris |
|---|---|
| GW Leo | $T_0 = BJD_{TDB}\ 2452721.55682\ (^{+0.00010}_{-0.00010}) + 0.33614880\ (^{+0.00004296}_{-0.00005912}) \times E$ days |
| QT Boo | $T_0 = BJD_{TDB}\ 2451402.54158\ (^{+0.00213}_{-0.00316}) + 0.31906397\ (^{+0.00000018}_{-0.00000013}) \times E$ days |

According to the results, the changing rate of the period was measured as $\frac{dp}{dt} = -6.21 \times 10^{-3}\ days\ yr^{-1}$ for GW Leo, and $\frac{dp}{dt} = -4.72 \times 10^{-3}\ days\ yr^{-1}$ for QT Boo. This downward linear trend is evident in both cases. However, to perform a precise study of period variations, it needs more observation in a good period time.

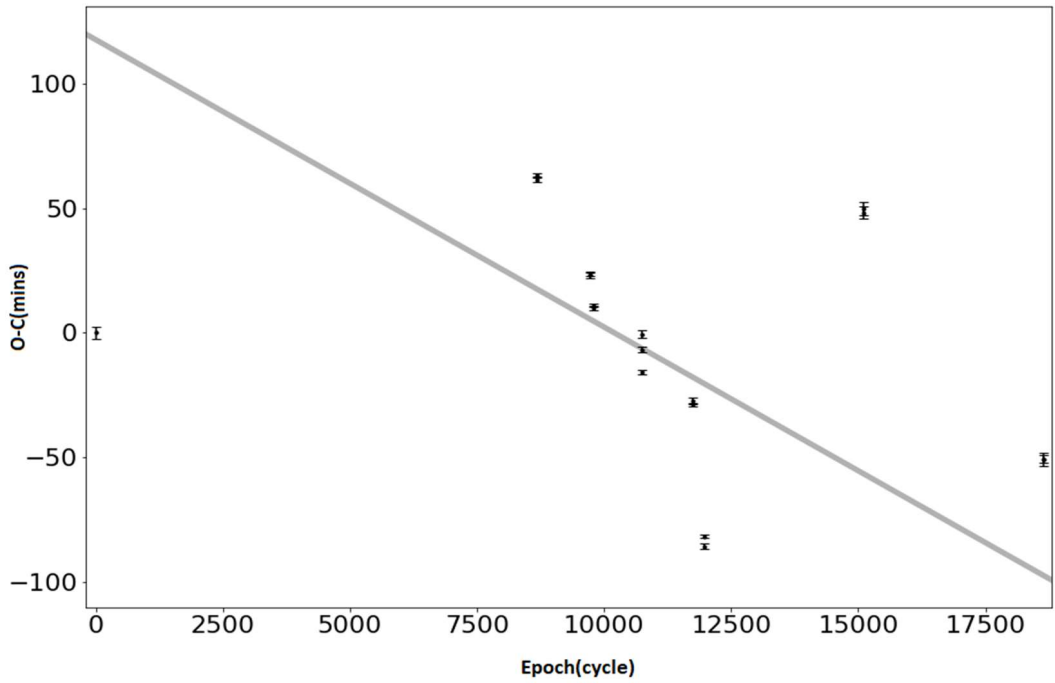

**Figure 3.** The O-C diagram of GW Leo with the derived line fitted on the data.

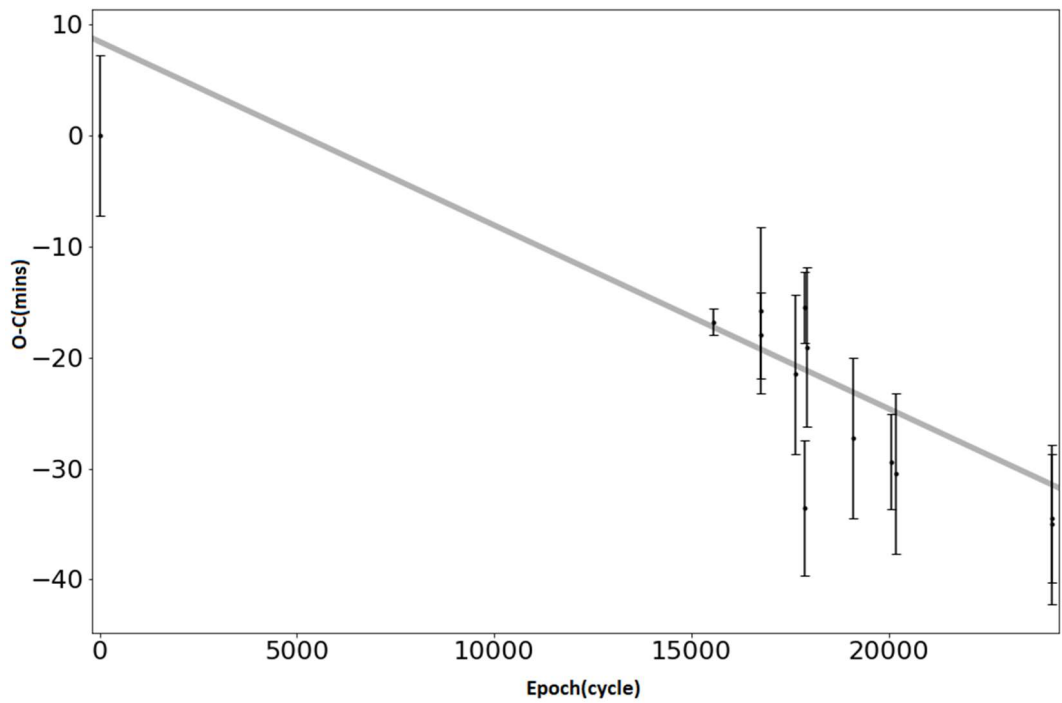

**Figure 4.** The O-C diagram of QT Boo with the derived line fitted on the data.

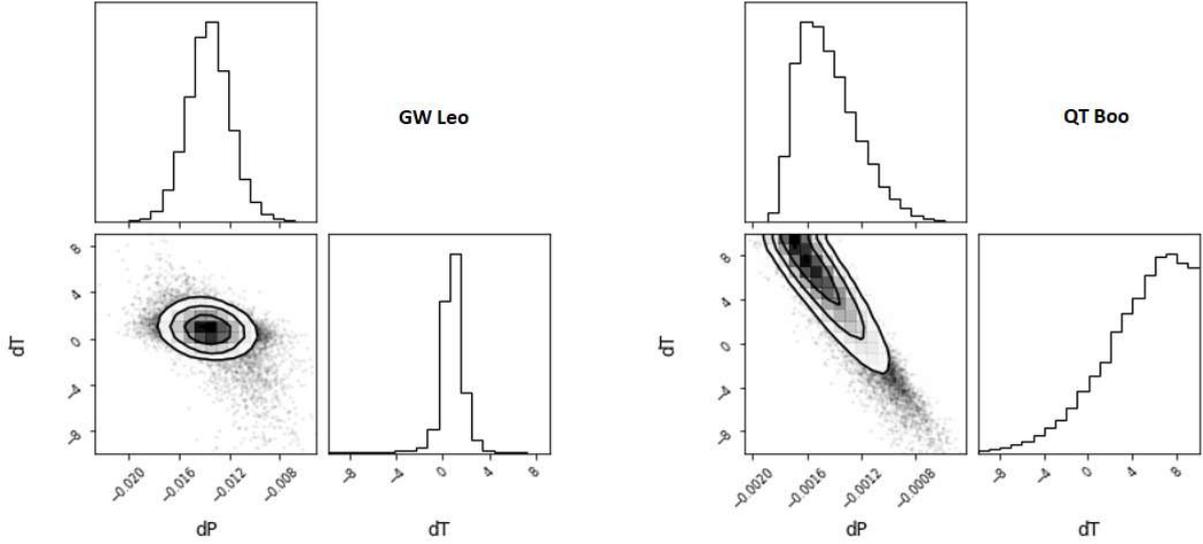

**Figure 5.** Corner plot showing the two-dimensional probability distributions of dT and dP and the histograms posterior probability distribution of both.

### 4. Light curve analysis of GW Leo

To determine the physical parameters of GW Leo, we analyzed the light curves using the binary star model of Wilson & Devinney (W-D) (Prša and Zwitter 2005) in mode 3.

The parameters considered to be fixed in the (W-D) program are the gravity darkening exponents $g_c = g_h = 0.32$ (Lucy 1967), the bolometric albedo coefficients $A_c = A_h = 0.50$ (Rucinski 1973), the effective temperature of the cooler component drove from Gaia DR2, and linear limb darkening coefficients adopted from tables published by Van Hamme (1993).

Since there is no spectroscopically determined mass ratio ($q$), we employed $q$-search to examine a series of values from 0.1 to 1.5. Orbital inclination ($i$), the surface temperature of the hotter star ($T_h$), relative monochromatic luminosity of the cooler star ($L_c$), and omega-potentials of the components ($\Omega_c = \Omega_h$) are considered as adjustable values through the W-D code's evaluation. The sum of the squared residuals $\sum W(O-C)^2$ for the tested mass ratios ($q$) is plotted in Figure 6. We selected $q = 0.881 \pm 0.030$ with minimal residuals as the initial input in the differential-correction stage of the W-D code. Derived parameters from the light curve solutions for GW Leo are listed in Table 6.

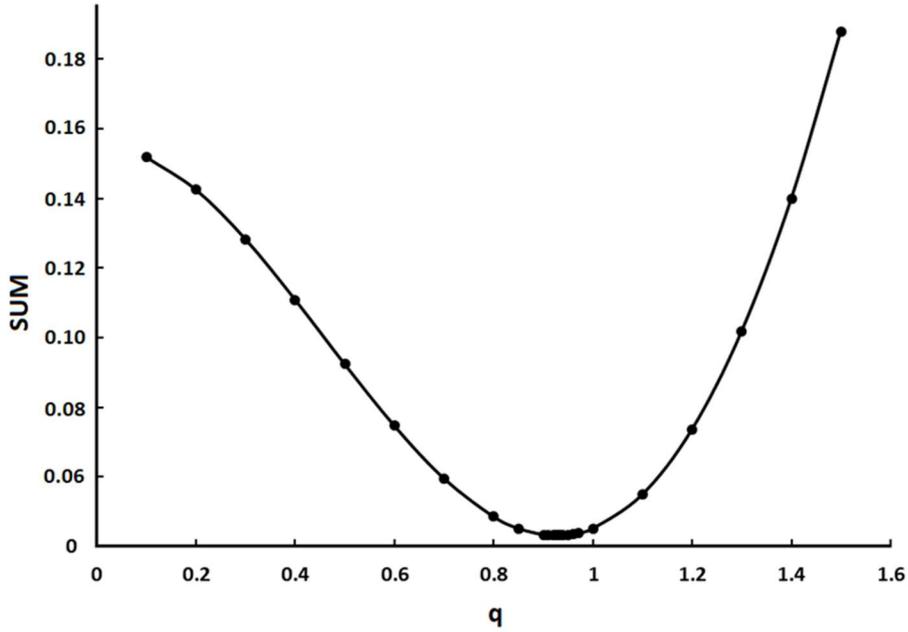

**Figure 6.** Sum of the squared residuals as a function of the mass ratio.

Table 6. Photometric solutions of GW Leo.

| Parameter | Results |
|---|---|
| $T_c\ (K)$ | 5832 |
| $T_h\ (K)$ | 6315(41) |
| $\Omega_c = \Omega_h$ | 3.54(43) |
| $i\ (deg)$ | 54.06(66) |
| $q$ | 0.881(3) |
| $l_c/l_{tot}$ | 0.44(7) |
| $l_h/l_{tot}$ | 0.56 |
| $A_c = A_h$ | 0.50 |
| $g_c = g_h$ | 0.32 |
| $f\ (\%)$ | 3 |
| $r_c(back)$ | 0.417 |
| $r_c(side)$ | 0.386 |
| $r_c(pole)$ | 0.367 |
| $r_h(back)$ | 0.394 |
| $r_h(side)$ | 0.363 |
| $r_h(pole)$ | 0.346 |
| $r_h(mean)$ | 0.389(17) |
| $r_h(mean)$ | 0.367(18) |
| $Colatitude_{spot}\ (deg)$ | 111 |
| $Longitude_{spot}\ (deg)$ | 116(6) |
| $Radius_{spot}\ (deg)$ | 27(11) |
| $T_{spot}/T_{star}$ | 0.80(4) |
| $Phase\ Shift$ | -0.035(2) |

Generally, the asymmetric light curves of W UMa type systems may demonstrate the presence of starspot on one or both stars in the system. We produced starspot-free, symmetric model light curves, which were fitted to the noticed brighter maximum of each light curve for each epoch. Turn to form, the model light curves did not fit the observed light curves, which are markedly asymmetric. Due to the deviations obvious on the maximums of the light curves known as the O'Connell effect (1951), a starspot was placed on the cooler component.

The mass of the cooler star was obtained from a study by Eker et al. (2018), and the mass of the hotter star can be calculated via the mass ratio equation. The absolute parameters of GW Leo are measured and given in Table 7. We provided an estimation of components mass through the photometry; For highly precise results and less amount uncertainties, spectral data are required.

Table 7. Estimated absolute parameters of GW Leo.

| Parameters | Cooler star | Hotter star |
|---|---|---|
| $Mass\ (M_\odot)$ | 1.077 | 0.949(21) |
| $Radius\ (R_\odot)$ | 1.001(45) | 0.944(48) |
| $Luminosity\ (L_\odot)$ | 1.04(52) | 1.27(56) |
| $M_{bol}$ | 4.70(15) | 4.48(16) |
| $log\ g$ | 4.469(38) | 4.465(39) |
| $a/R_\odot$ | 2.573(4) | |

Regarding to the quantities obtained for the absolute parameters, we computed the distances of GW Leo. The values of $m_v = 13.13$ and $M_v = 4.76$ were calculated for the cooler component. We used $BC_1 = -0.06$ according to Eker et al. (2018). Hence, we measured the distance to the binary system by inserting certain parameters in the equation below,

$$d_{(pc)} = 10^{(\frac{m_{pri}-M_{pri}+5-A_V}{5})} \quad (1)$$

which yielded $465.58 \pm 23$ pc using $A_v = 0.03$ (Schlafly and Finkbeiner, 2011).

The 3D view of this binary system is shown in Figure 7.

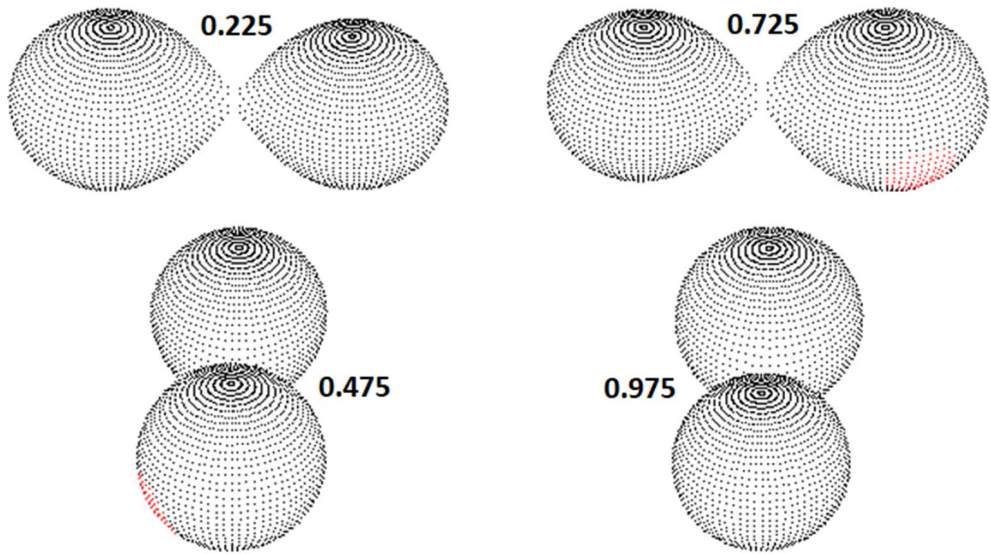

Figure 7. The 3D view of the stars.

## 5. Results and conclusion

The photometric observations of two W UMa binary systems GW Leo and QT Boo were carried out in two observatories and using a *V* filter.

A new ephemeris was suggested to determine the times of primary minima for both contact binary GW Leo and QT Boo. Besides the O-C diagrams for two binary systems, the long-term decreases are noticeable. The orbital periods are decreasing at a rate of $\frac{dp}{dt} = -6.21 \times 10^{-3} \, days \, yr^{-1}$ for GW Leo, and $\frac{dp}{dt} = -4.72 \times 10^{-3} \, days \, yr^{-1}$ for QT Boo. This decrease in contact binaries may be interpreted by either the mass transfer from the more massive component to the less massive component or mass and angular momentum loss from the binary system (Yang et al. 2007).

The first photometric solutions of the short period binary system GW Leo were done by analyzing its light curve. As a result, it is found that GW Leo is a contact binary with a mass ratio of $q = 0.881 \pm 0.030$, a fillout factor of $f$ = 3%, and an inclination of $i = 54.060 \pm 0.066 \, deg$. As indicated by the light curve solution, a cool starspot is placed on the cooler component. The various temperatures of components for this binary system are close to 500 k, so the components are relied upon to have a magnetic activity that may cause the observed noticed unbalanced light curves of GW Leo. This recommendation for a magnetic activity can be check with future spectrometric observations and investigation of period varieties with more occasions of minima.

We estimated of absolute parameters related to both components of the GW Leo. Mass, radius, and luminosity of the system were obtained using the binary systems modeling of the Wilson-Devinney. According to the estimated absolute parameters, we measured the distance as $465.58 \pm 23$ pc. This value is relatively close to the quantity obtained by the Gaia DR2 as $483.56 \pm 10$ pc.

The components' positions of GW Leo are plotted in the Hertzsprung-Russell (H-R) diagram (Figure 8), which seems both of them near the main-sequence. With one of the components near the ZAMS and the other components in the middle of the main-sequence.

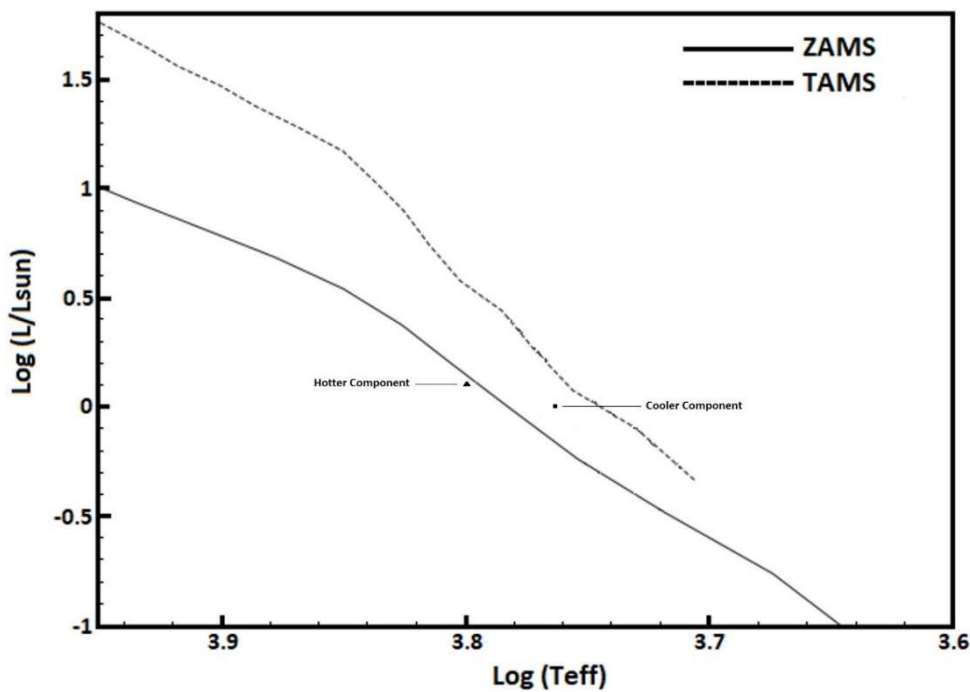

**Figure 8.** Position of GW Leo on the H-R diagram, in which the theoretical ZAMS and TAMS curves are indicated.

## Acknowledgments

This manuscript was prepared by the International Occultation Timing Association Middle East section (IOTA/ME) as a scientific sponsor in the form of the APTO project, which has been carried out between several observatories in Iran during 2019 and 2020. We would like to thanks Dr. Mohammad Fadaeian for his cooperation and to provide the Payame Noor Observatory of Ardabil University for this project. Also, thanks to PegahSadat MirshafieKhozani to make some corrections in the text. Furthermore, many thanks to Fatemeh Davoudi for her scientific cooperation.